\documentstyle[prl,aps,psfig]{revtex} 

\begin{document}


\title{From Generalized Synchrony to Topological Decoherence:\\
Emergent Sets in Coupled Chaotic Systems}
\author{Ernest Barreto$^*$, Paul So$^*$, Bruce J. Gluckman$^*$, and Steven J. Schiff$^\dag$}
\address{George Mason University, Fairfax Virginia 22030 \\
$^*$ Department of Physics and Astronomy and the Krasnow Institute \\
$^\dag$ Department of Psychology and the Krasnow Institute}
\date{\today}
\maketitle
\begin{abstract}
We consider the evolution of the unstable periodic orbit structure of coupled chaotic systems. This involves the creation of a complicated set outside of the synchronization manifold (the {\it emergent set}). We quantitatively identify a critical transition point in its development (the {\it decoherence} transition). For asymmetric systems we also describe a migration of unstable periodic orbits that is of central importance in understanding these systems. Our framework provides an experimentally measurable transition, even in situations where previously described bifurcation structures are inapplicable.
\end{abstract}



The idea that several subsystems, when interacting nonlinearly, collectively give rise to novel dynamics that are not obviously attributable to the individual component parts has been termed {\it emergence} \cite{emergence}. In this Letter we investigate such novel dynamics in systems of coupled chaotic maps, with an emphasis on systems of dissimilar components. When synchronized, the time evolution occurs on a restricted manifold (called the synchronization manifold) embedded in the full state space. As the degree of coupling is decreased to zero, the system gradually evolves into a completely unsynchronized state in which all the degrees of freedom of the individual component maps are realized. At each extreme, the dynamics can be understood in terms of the components. In between, however, the situation is more complicated.

Various transitions in this desynchronization process have been described in the literature \cite{IS,bubbling,riddle,blowout,astakov,maistrenko,zaks,just,gs,dgs}. Much of this earlier work depends on an invariant manifold that persists under decreased coupling due to system symmetries. Our approach is novel in that it does not refer to invariant manifolds or symmetries. Instead, we analyze the evolution of the periodic orbit structure as the coupling is decreased and synchronization breaks down. Our formalism is therefore applicable to a much larger class of coupled systems, in particular, those consisting of {\it dissimilar} components. We report two main results. We describe the creation and evolution of a complicated set that develops outside of the synchronization manifold (the {\it emergent set}), and we quantitatively identify a critical transition point in its development (the {\it decoherence} transition). For asymmetric systems we also describe a migration of unstable periodic orbits that is of central importance in understanding these systems. Our framework is advantageous because it provides an experimentally measurable transition in situations where previously described bifurcation structures are inapplicable.

Previous work has focused on the invariant dynamics in the synchronization manifold $\mathcal{M}$, which can easily be identified in coupled systems with symmetry (such as when two identical sub-systems are coupled together). On $\mathcal{M}$, the components evolve identically, and are said to exhibit identical synchrony \cite{IS}. As the coupling decreases from a fully synchronized state, a bubbling bifurcation \cite{bubbling} occurs when an orbit within $\mathcal{M}$ (usually of low period \cite{lowperiod}) loses transverse stability. In the presence of noise or {\it small} asymmetries, a typical trajectory quickly approaches and spends a long time in the vicinity of $\mathcal{M}$, but makes occasional excursions. (If an appropriate attractor exists outside the synchronization manifold, this transition leads to the creation of riddled basins \cite{riddle}.) As the coupling is further decreased, the blowout bifurcation \cite{blowout} is observed when $\mathcal{M}$ itself becomes  transversely unstable (on average). More recent work has described bifurcations that lead to the creation of periodic orbits off the synchronization manifold \cite{astakov}; these may lead to the creation of chaotic attractors external to $\mathcal{M}$ \cite{maistrenko}. Also, imperfect phase synchrony has been analyzed recently in terms of unstable periodic orbits \cite {zaks}, and synchrony transitions have been investigated in coupled lattices of identical maps \cite{just}.

The concept of (differentiable) generalized synchrony (GS) \cite{gs,dgs} extends these ideas. GS relaxes the condition that the state variables evolve identically, and only requires that they be functionally related. As the coupling is reduced and GS breaks down, however, this function may become extremely complicated, and the identification of bubbling-type or blowout-type bifurcations is especially problematic. Thus, a more general description of the desynchronization process is needed.

In the present work we extend the above description by considering the evolution of the unstable periodic orbits (UPOs) as the coupling is varied. We use the following two-dimensional, unidirectionally coupled system \cite{josic}:
\begin{equation}
\left\{
\begin{array}{ccl}
x & \rightarrow & f(x) \\
y & \rightarrow & cf(x) + (1-c)g(y),
\end{array}
\right.
\label{general_system}
\end{equation}
where $f$ and $g$ are chaotic maps and $c$ is a scalar that describes the coupling. We emphasize that $f$ and $g$ need not be similar, and may be of any dimension. For example, we have also studied a four-dimensional system in which $f$ and $g$ are H\'{e}non maps with different parameters. For ease of presentation we restrict discussion to one-dimensional $f$ and $g$. Systems such as Equation (\ref{general_system}) are known in the mathematical literature as skew products or extensions; ours is constructed such that for $c=1$, the $x$ and $y$ dynamics are identical and synchronized, whereas for $c=0$, the $x$ and $y$ dynamics are completely independent. We investigate this system as $c$ is decreased from $1$ to $0$.

The simplest case is when $f=g$, for which the synchronization manifold $\mathcal{M}$ (i.e. the line $x=y$) is invariant and attracting at $c=1$. The bubbling bifurcation occurs when an orbit in $\mathcal{M}$ loses transverse stability, typically via a period-doubling (pitchfork) bifurcation. This leads to the creation of new orbits outside of $\mathcal{M}$. As the coupling is further reduced, more and more periodic orbits embedded in $\mathcal{M}$ lose their transverse stability in a similar fashion \cite{additionalnote}, leading to the creation of additional orbits. As this process proceeds, the external UPOs simultaneously undergo period-doubling cascades to chaos, thus creating even more new orbits. We call the set of new orbits created in this fashion the {\it emergent set}.

In the more general case $f \neq g$, $x=y$ is by construction invariant and attracting for $c=1$. Upon decreasing $c$, $x=y$ is no longer invariant, and we observe that the UPOs migrate and spread out as shown in Figure \ref{upospread} for coupled quadratic maps. As the coupling is decreased, we first observe transverse Cantor-like structure, followed by a ``fattening" of the striations as the Lyapunov dimension of the attractor increases to $2.0$\cite{beck}. We have also observed similar UPO migration in the invertible case of coupled H\'{e}non maps. It is remarkable that this UPO migration appears to occur well before any orbit loses its transverse stability. In fact, we observe a large range of $c$ over which, despite the apparent structure (ultimately two-dimensional), the periodic orbits migrate but are still transversely stable and one-to-one in the following sense: if the driver dynamics is fixed onto any one of its intrinsic period $p$ orbits, then the limiting $y$ dynamics is an attracting orbit {\it of the same period}.

Let $U$ be the set of unstable periodic orbits on the line $x=y$ when $c=1$. The number of orbits in $U$ is determined by the driver and remains constant for all $c$ because of the unidirectional coupling. For $f=g$, they remain fixed in place, but for $f \neq g$, they migrate as described above. As $c$ is decreased from $1$, the orbits' stability properties evolve, but they remain transversely attracting until a bubbling-type bifurcation is encountered. (We extend the concept of bubbling to the asymmetric case $f \neq g$ by defining it as the point where the first orbit in $U$ loses stability \cite{firstbifnote}.) As the coupling is further decreased, more and more orbits bifurcate and create orbits outside of $U$, and the above mechanism for the creation of the emergent set applies. Because of their migration, however, the orbits of $U$ become intermingled among those of the emergent set.

We wish to view system (\ref{general_system}) as an infinite collection of subsystems defined as follows. First, enumerate the periodic orbits of $f$ (the driver), assigning each an index $i=1, 2, \dots$. Then subsystem $S_i$ is given by Equation (\ref{general_system}), but with the driver dynamics $f$ locked on orbit $i$. The bifurcations described above correspond to bifurcations in the $y$ components of these subsystems. Indeed, each subsystem $S_i$ exhibits a complete bifurcation structure in $y$ as $c$ is varied from $1$ to $0$.

To quantify the discussion, let $N_{xy}(p)$ denote the number of period $p$ orbits of system (\ref{general_system}). Also, let $N_f(p)$ denote the number of period $p$ solutions of $f^p(x) -x = 0$ alone; $N_g(p)$ is defined analogously. These quantities contain contributions from all periodic orbits of period $q$, where $q$ is an integer factor of $p$. For $c=1$, the system exhibits identical synchrony and $N_{xy}(p) = N_f(p)$. In contrast, when $c=0$, the system is fully decoupled into independent systems, and $N_{xy}(p) = N_f(p)N_g(p)$. (Note that $N_{xy}(p)$ may achieve its maximum at $c=0$ or at intermediate values of $c$, depending on the nature of $g$.)

Our goal is to elucidate how this change in the unstable periodic orbit structure proceeds as $c$ is varied. To this end, we consider the topological entropy $h$ \cite{joeri}; for large $p$, the number of periodic orbits of period $p$ in a chaotic set increases exponentially with $p$ as $N(p) \simeq e^{hp}$. Thus, the topological entropy of the coupled system is $h_{xy} = \lim_{p \rightarrow \infty} ln N_{xy}(p)/p$, and similarly, the topological entropy of the driver is $h_f = \lim_{p \rightarrow \infty} ln N_f(p)/p$.
 
Let $N_e(p)$ be the number of periodic orbits of period $p$ that are not in $U$. These orbits reside in the emergent set and are created by the bifurcations described above. Thus
$ N_e(p) = \sum_{i=1}^{n_b} N_i(p), $
where the summation is over the number $n_b$ of subsystems that have bifurcated. $N_i(p)$ is the number of periodic orbits of period $p$ not in $U$ that are associated with a particular subsystem $S_i$ in the summation. The topological entropy of the emergent set is therefore $h_e = lim_{p \rightarrow \infty} N_e(p)/p$ \cite{emergentnote}.

We define the state of {\it topological coherence} for system (\ref{general_system}) as the condition $h_{xy}=h_f$. In this state, the topological entropy of the system is determined by the driver. In order for topological coherence to be destroyed, the topological entropy of the full system $h_{xy}$ must exceed $h_f$. This occurs when the emergent set becomes sufficiently complex. In the present case, 
$ h_{xy} = \lim_{p \rightarrow \infty} \frac{1}{p} ln \left( N_f(p) + N_e(p) \right) = max(h_f, h_e)$. 
There is therefore a critical value $c_d$ of coupling where $h_e$ first exceeds $h_f$ and topological coherence is lost. We call this the {\it decoherence transition}. For the special case $f=g$, we find that this typically occurs between the bubbling $c_{bu}$ and the blowout $c_{bo}$ bifurcations.

We now address the question of how the decoherence transition may be measured from experimental data. First, we observe that in the symmetric noise-free case ($f=g$), trajectories collapse onto $\mathcal{M}$ and remain there until the blowout bifurcation. Thus, estimates of the decoherence transition based on measured data will not reflect the contribution of the emergent set. However, this case is exceptional. In the more general asymmetric case ($f \neq g$), the orbits of $U$ migrate and become intermingled with those of the emergent set. Because of this, typical trajectories do not necessarily remain near $U$; instead, the observed attractor incorporates {\it parts} of the emergent set. How much of the emergent set is incorporated depends on the degree of asymmetry and the coupling. By using trajectory data, an {\it effective decoherence transition} can be measured which indicates how much the emergent set actually influences the observed dynamics. It is precisely this effective transition that is most relevant to the observed dynamics of the system and hence is most relevant to experimental situations. Below we describe an efficient method for estimating the effective decoherence transition that is based on actual trajectory information. 

We use the methods of Ref. \cite{joeri}. These authors define an average $n$-step stretching rate as follows. Let $\lambda^{(n)}_i$ denote the square root of the largest eigenvalue of $[{\bf J}^n({\bf x}_i)]^T{\bf J}^n({\bf x}_i)$ for some initial condition ${\bf x}_i$, where ${\bf J}$ is the Jacobian of the system in Equation (\ref{general_system}). Then form the following average quantity over $m$ initial conditions chosen with respect to the natural measure:
$ln \langle 1d \rangle_{n,m} = ln \left(
\sum_{i=1}^{m} \lambda_{i}^{(n)} /m \right)/n$.
For hyperbolic systems with one stretching direction, it can be shown \cite{joeri} that $h_1 = \lim_{n \rightarrow \infty} \lim_{m \rightarrow \infty} ln \langle 1d \rangle_{n,m}$ is the topological entropy of the system \cite{lyapnote}. Numerically, $h_1$ can be obtained by measuring the scaling of $ln (\sum_{i=1}^{m} \lambda_{i}^{(n)})$ with $n$, for a sufficiently large $m$.

Since Equation (\ref{general_system}) can exhibit two expanding directions for certain parameter values, we also measure the two dimensional average stretching rate: 
$ln \langle 2d \rangle_{n,m} = ln \left(
\sum_{i=1}^{m}
(\lambda_1 \lambda_2)_{i}^{(n)} /m \right)/n$,
where $(\lambda_1 \lambda_2)^{(n)}_i$ represents the square root of the product of the two largest eigenvalues of $[{\bf J}^n({\bf x}_i)]^T{\bf J}^n({\bf x}_i)$. The topological entropy when there are two expanding directions is given by $h_2 = \lim_{n \rightarrow \infty} \lim_{m \rightarrow \infty} ln \langle 2d \rangle_{n,m}$.

These quantities enable us to calculate the topological entropy of the full system as the coupling parameter varies and traverses regions with one and two stretching directions: $h_{xy} = max(h_1, h_2)$. For the examples considered below, we have $h_1=h_f$ for the entire coupling range of interest. Thus, the {\it effective} decoherence transition occurs when $h_2$ first exceeds $h_1=h_f$. (In higher dimensional systems, these methods are practical if the number of unstable directions is low; otherwise, it may be difficult to accurately calculate these quantities from limited data \cite{EkRu}.)

We apply these methods to a system of coupled quadratic maps. We take $f(x) = 1.7 - x^2$, $g(y) = a_g - y^2$, and consider the cases $ a_g = 2.0$, $1.7$, and $1.5$. (We have also used two H\'{e}non maps coupled as in Eq.~\ref{general_system}, and the results are qualitatively the same.) Figure \ref{quadraticgraph} shows $max(h_1,h_2)$ for these cases. In all cases, $h_1$ is equal to the topological entropy of the driver dynamics $h_f$ (for $a_g=2.0$, $h_1 > h_f$ only for $c<0.1$, not shown). The effective decoherence transition occurs when $h_2$ exceeds $h_1$, as indicated by arrows.

Finally, we illustrate the influence of the emergent set for the symmetric case $a_f=a_g$ with noise. As $c$ is decreased from bubbling, the dynamics will make occasional excursions from $\mathcal{M}$. We expect progressively longer transient times outside of $\mathcal{M}$ due to the increasing complexity of the emergent set. To illustrate, we consider the symmetric case $a_f=a_g=1.7$ and plot in Figure \ref{burstgraph} the average duration of a burst versus coupling value. We define a burst as an excursion of at least $10$ iterations beyond a small distance $\delta=0.05$ from the line $x=y$; the end is when the orbit falls back within this distance. We measure iterations per burst over trajectories of $10^6$ iterations, then average over $100$ such realizations. For each noise level (uniformly distributed noise within a given amplitude), there is a transition range of $c$, depending on $\delta$ and the magnitude of the noise, above which bursts are not observed. For very small noise, this transition is close to the blowout bifurcation; for larger noise, the transition shifts to higher values of coupling. Note that for $c$ values below their respective transitions, the various curves asymptote to a common curve, suggesting that the dynamics during the bursts is consistently influenced by the emergent set. As expected, the average duration of bursts increases with decreasing $c$.

Finally, we note that the mechanism for the creation of emergent sets outlined here leads one naturally to expect unstable dimension variability \cite{UDV} as a typical feature of emergent sets, and of coupled systems in general.

In conclusion, we emphasize that the emergent set framework developed here is quite general and applies to coupled systems of non-identical elements where previously studied bifurcation structures may be problematic or inappropriate. Furthermore, the effective decoherence transition can be estimated in such systems from experimental data.

We thank C. Grebogi, E. Ott, B. Hunt, and J. A. Yorke. This work was supported by NSF (IBN 9727739, E.B. and P.S.) and NIH (7K02MH01493, S.J.S.; 2R01MH50006, S.J.S., P.S., and B.J.G.). Figure \ref{upospread} was generated using {\tt DYNAMICS} \cite{dynamics}.

\begin{figure}
\caption{Magnifications of the attractor for Equation (\ref{general_system}) with $f(x) = 1.7 - x^2$, $g(y)=1.5 - y^2$, and $c=0.45$ (inset, $c=0.9$). Superimposed on both views are periodic orbits ($+$) of periods $1$ to $20$. At $c=1$, these orbits lie on the diagonal $x=y$, but as $c$ is decreased, they migrate as shown here. The Lyapunov dimension is $2.0$ (inset, $1.25$), and the orbits in both cases are transversely stable and one-to-one in the sense described in the text.}
\label{upospread}
\end{figure}

\begin{figure}
\caption{Topological entropy estimate $max(h_1, h_2)$ for the system in Equation (\ref{general_system}) with $f(x) = 1.7 - x^2$, $g(y) = a_g - y^2$, and cases $a_g = 2.0$, $1.7$, and $1.5$. $h_1$ in all cases equals $h_f$, the topological entropy of the driver, for the range of coupling shown. The effective decoherence transition occurs when $h_2$ first exceeds $h_1$ (arrows). (For the symmetric case $a_g=a_f=1.7$, bubbling occurs at $c_{bu} \simeq 0.442$ and blowout at $c_{bo} \simeq 0.352$.)}
\label{quadraticgraph}
\end{figure}

\begin{figure}
\caption{
The duration of bursts away from the synchronization manifold for Equation (\ref{general_system}) with $f(x) = 1.7 - x^2$ and $g(y) = 1.7 - y^2$. The threshold value of $c$ for observing such bursts increases with larger noise. The curves asymptote to a common curve for coupling less than this threshold, indicating that the trajectory during the bursts is strongly influenced by the emergent set.
}
\label{burstgraph}
\end{figure}


\begin{references}
\bibitem{additionalnote}
Additional orbits external to $\mathcal{M}$ may be created by saddle-node bifurcation; these may then exchange stability with their corresponding orbits in $\mathcal{M}$ via a transcritical bifurcation.
\bibitem{firstbifnote} In all examples studied, the first such bifurcation is a period-doubling bifurcation.
\bibitem{lyapnote}
If the logarithm is taken before the average, the result is the Lyapunov exponent. The procedure used here has contributions from other structure in the finite-time Lypunov exponent distribution.
\end{references}
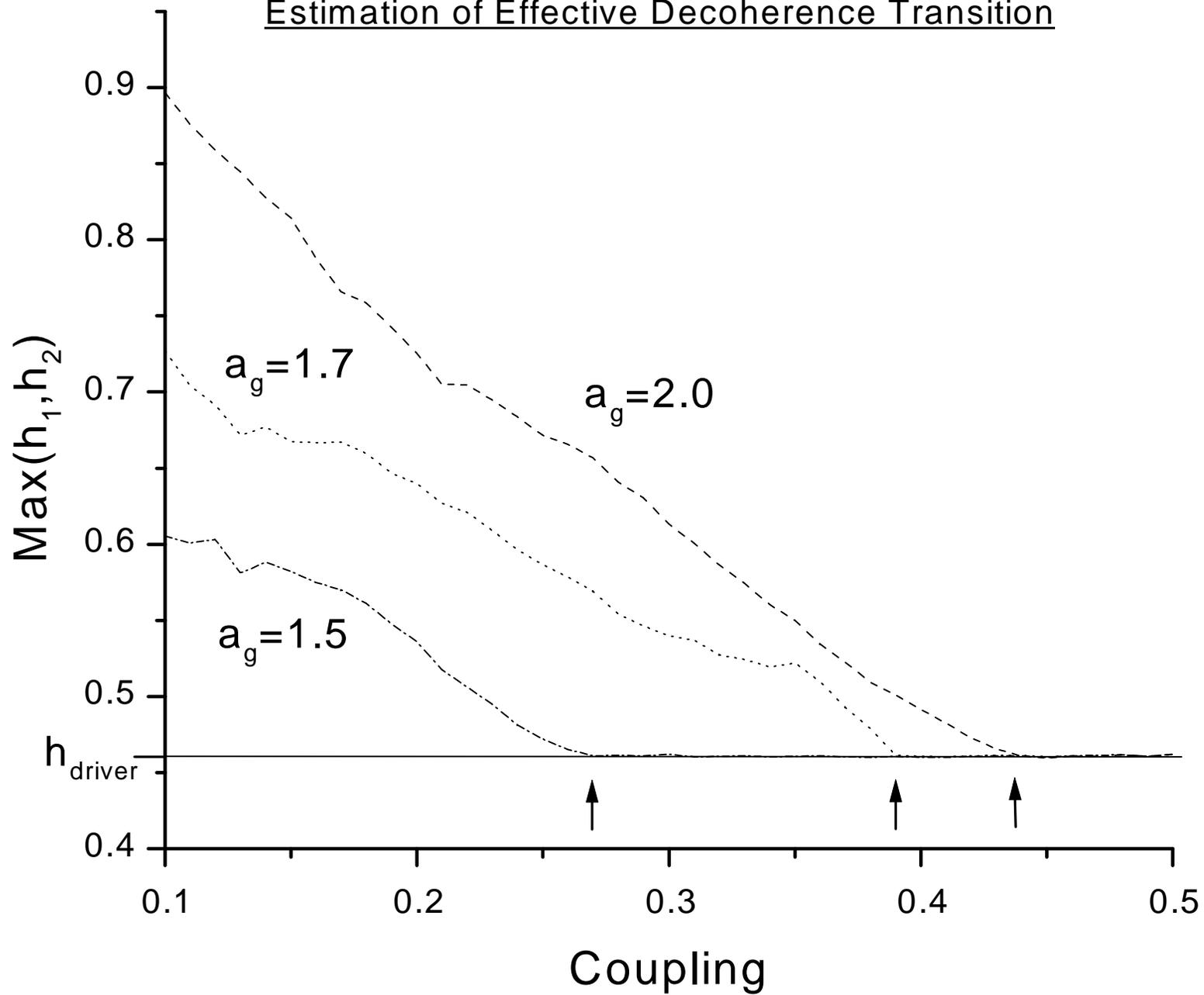
\end{document}